\documentclass[review]{elsarticle}
\usepackage{amssymb}
\usepackage{amsmath}
\usepackage{tabularx}
\usepackage[english]{babel}
\usepackage{graphicx}
\usepackage{color}
\usepackage{multirow}
\usepackage{subfig} 
\usepackage{bm}
\usepackage{algorithm}
\usepackage{threeparttable}

\newtheorem{thm}{Theorem}
 
\newtheorem{propo}[thm]{Proposition}
\newdefinition{rmk}{Remark}
\newdefinition{define}{Definition} 
\newproof{pf}{Proof}

\journal{Elsevier}

\begin{document}

\begin{frontmatter}

\title{Positivity-preserving method for multi-resolution simulations of compressible flows}

\author{Shucheng Pan} \ead{shucheng.pan@tum.de}
\author{Xiangyu Hu} \ead{xiangyu.hu@tum.de}
\author{Nikolaus. A. Adams} \ead{nikolaus.adams@tum.de}
\address{Lehrstuhl f\"{u}r Aerodynamik und Str\"{o}mungsmechanik, Technische Universit\"{a}t
M\"{u}nchen, 85748 Garching, Germany}
\begin{abstract}
We present a positivity-preserving method for multi-resolution simulations of compressible flows involving extreme conditions such as near vacuum and strong discontinuities. The novelty of this work is due to two aspects. First we extend the positivity-preserving flux limiter of (Hu et al., J Comput Phys 242, 2013) to the multi-resolution framework by modifying the prediction operator, based on the same limiting concept. Second, we develop a positivity-preserving local time stepping scheme for adaptive time marching. Instead of using fixed hierarchical time steps, the local time stepping scheme dynamically adjusts the time steps of all multi-resolution levels to maintain positivity. The method is validated and its capabilities are demonstrated by a range of test cases.
\end{abstract}

\end{frontmatter}


\section{\label{sec:intro}Introduction}
High-order conservative schemes, such as the essentially non-oscillatory (ENO) \cite{harten1983high} and the weighted essentially non-oscillatory (WENO) schemes \cite{jiang1996efficient}, are widely used in simulations of compressible flows, as they have the capability to resolve simultaneously small flow structures and shock waves. Unlike first-order schemes which maintain positive density and pressure, such conservative high-order schemes may develop oscillatory spurious solutions on the level of the truncation error and thus may produce negative density or pressure for flows near vacuum and strong discontinues. Although simply clipping or using non-conservative formulations can prevent such failure, this may result in wrong shock locations or nonlinear numerical instability \cite{einfeldt1991godunov}. To impose the positivity-preserving property to high-order conservative schemes, Zhang et al. \cite{zhang2010positivity, zhang2012positivity} have developed a positivity-preserving flux limiter which is suitable for discontinuous Galerkin methods and WENO schemes and is based on Legendre-Gauss-Lobatto quadrature. This limiter has been successfully applied for the simulation of magnetohydrodynamics \cite{zhang2011positivity, cheng2013positivity} and multi-material compressible flows \cite{cheng2014positivity, vilar2016positivity}. An alternative approach is proposed by Hu et al. \cite{hu2013positivity} who detect negative density/pressure locations \textit{a posteriori} and employ a convex combination of the high-order numerical flux and the first-order Lax-Friedrichs flux to satisfy a sufficient condition for preserving positivity. The main advantage of this limiter is that the time step constraint is less restrictive than for the method of Zhang et al. \cite{zhang2010positivity, zhang2012positivity}, and that it can be applied to any high-order conservative scheme \cite{kotov2013comparative} without deteriorating its formal accuracy. This simple positivity-preserving flux limiter has been extended to relativistic hydrodynamics \cite{radice2014high, wu2015high, porth2016black}.

Adaptive discretizations have become a powerful tool for simulations of complex compressible flows containing a broad range of temporal and spatial scales. Adaptive mesh refinement (AMR) \cite{berger1984adaptive, berger1989local, macneice2000paramesh} and wavelet-based multi-resolution methods (MR) \cite{harten1994adaptive, harten1995multiresolution, roussel2003conservative} employ variable grid resolution levels according to a local error estimate. Compared to AMR the MR method typically achieves higher memory compression rates \cite{deiterding2009adaptive} and allows for a more rigorous regularity analysis \cite{harten1995multiresolution, muller2012adaptive}. Local scale-dependent time-stepping schemes (LTS) are introduced to achieve additional speed-up during time marching \cite{osher1983numerical, dawson2001high}. By combining MR and LTS, space-time adaptive methods \cite{domingues2008adaptive, domingues2009space} offer considerably improved efficiency. Such methods can be further improved by formulating the adaptive algorithm for efficient parallel execution \cite{hejazialhosseini2010high,han2011wavelet,han2014adaptive}. High-order finite-difference WENO schemes \cite{jiang1996efficient}, in conjunction with a space-time MR framework \cite{sjogreen2004multiresolution, burger2007adaptive, han2011wavelet, han2014adaptive} enable efficient high-resolution simulations of compressible flows. In this case, however, a straightforward application of flux limiters \cite{zhang2012positivity, hu2013positivity} developed for uniform grids in the MR framework 
is not sufficient to achieve the overall positivity-preserving property.
One issue is that the prediction operator \cite{harten1995multiresolution}, which relies on high-order interpolation, may produce negative density or pressure during mesh refinement. The other is that LTS \cite{domingues2008adaptive} with time steps fixed during a full cycle and with the conservative flux correction applied at cell faces shared by different levels, may also lead to positivity violation. This latter issue, to our knowledge, has not been addressed yet by  methods in the literature. 

The objective of the present paper is to develop a simple positivity-preserving method for MR discretization for compressible-flow evolution involving vacuum and strong discontinuities. Adaptation method, pyramid data structure and parallel strategy are based on Ref. \cite{han2014adaptive}. The paper is organized as follows. Sec. \ref{sec:prelimi} gives a brief overview the employed high-order conservative schemes. In Sec. \ref{sec:num}, we discuss how to achieve the positivity-preserving property in the MR and LTS framework. Sec. \ref{sec:example} is dedicated to assessing the capabilities of the present method. Concluding remarks are given in Sec. \ref{sec:con}.
\section{Preliminaries \label{sec:prelimi}}
The governing equations of an invisicid compressible flow are the one-dimensional Euler equations
\begin{equation}\label{govern}
\frac{\partial \mathbf{U}}{\partial t}+ \frac{\mathbf{F}(\mathbf{U})}{\partial x} =0,
\end{equation}
where $\mathbf{U} = (\rho, \rho u, E)^T$, in which $\rho$, $u$, and $E$ are the density, velocity and the total energy with relation $E = \rho e + \rho u^2/2$, with $e$ being the specific internal energy. The  flux function is $\mathbf{F} = \left[ \rho u, \rho u^2 +p, (E+p) u \right]^T $ where $p$ is the pressure. To close the governing equations, the ideal-gas equation of state $p=(\gamma-1)\rho e$ is used to describe the thermodynamic properties of the materials, where $\gamma$ is the ratio of specific heats.

On a uniform 1D grid, Eq. (\ref{govern}) discretized with a $k$th-order conservative scheme and the explicit Euler time marching scheme is

\begin{equation}\label{governdis}
\mathbf{U}^{n+1}_{i} = \mathbf{U}^{n}_{i} + \lambda \left( \hat{\mathbf{F}}_{i-1/2} -\hat{\mathbf{F}}_{i+1/2} \right),
\end{equation}
where $\mathbf{U}^n_i$ and $\mathbf{U}^{n+1}_i$ are the cell averaged conservative variables of cell $\left[ x_{i-1/2},x_{i+1/2} \right]$. The superscript $n$ stands for the time step and $i$ for the cell index. The numerical flux $\hat{\mathbf{F}}_{i\pm 1/2} = \mathbf{F}_{i\pm 1/2} + \mathbf{O}(\Delta x^{k+1})$ depends on $\mathbf{U}_{i \pm 1/2}$ reconstructed from $\mathbf{U}_{j}$ or directly on a reconstruction from a primitive function for the flux. The parameter $\lambda = \Delta t/ \Delta x$, with $\Delta x$ and $\Delta t$ being the cell size and the time step size which satisfies the CFL condition,
\begin{equation}\label{CFL}
\Delta t = \frac{\mathrm{CFL}\, \Delta x}{\Vert |u|+c \Vert_{\infty}},
\end{equation}
where $c=\sqrt{\gamma p/\rho}$ is the sound speed and $\mathrm{CFL} \in (0,1)$, leading to

\begin{equation}
\lambda = \frac{\mathrm{CFL}}{\Vert|u|+c\Vert_{\infty}}.
\end{equation}
For more than one spatial dimensions, Eq. (\ref{governdis}) is extended appropriately dimension by dimension.
\subsection{Positivity-preserving flux limiter for high-order conservative schemes}\label{sec:pp}
In the following we revisit the positivity-preserving flux limiter \cite{hu2013positivity}. For a so-called finite difference WENO scheme \cite{jiang1996efficient}, the numerical fluxes $\hat{\mathbf{F}}_{i\pm 1/2}$ in Eq. (\ref{governdis}) are reconstructed at the cell-face $x_{i\pm1/2}$ and do not necessarily satisfy the positivity property. The flux limiter in Ref. \cite{hu2013positivity} maintains positivity by a convex combination of Lax-Friedrichs flux and $\hat{\mathbf{F}}_{i\pm 1/2}$.

Note that density function $\rho(\mathbf{U}) = \rho$ and pressure function $p(\mathbf{U}) = (\gamma-1)\left( E-\rho u^2/2 \right)$ are locally Lipschitz continuous and have the properties
\begin{equation}
\rho\left[ (1-\theta)\mathbf{U}_1 + \theta\mathbf{U}_2\right] = (1-\theta)\rho(\mathbf{U}_1) + \theta\rho(\mathbf{U}_2), \quad p\left[ (1-\theta)\mathbf{U}_1 + \theta\mathbf{U}_2\right] \geq (1-\theta)p(\mathbf{U}_1) + \theta p(\mathbf{U}_2)
\end{equation}
if $\rho(\mathbf{U}_1) \geqslant 0$, $\rho(\mathbf{U}_2) \geqslant 0$ and $\theta \in [0,1]$. Define the set of admissible states by

\begin{equation}
\mathbb{G} = \left\lbrace \mathbf{U} | \rho(\mathbf{U}), p(\mathbf{U}) \in \mathbb{R}^+ \right\rbrace 
\end{equation}
which is a convex set \cite{zhang2010positivity}. Given $\mathbf{U}^n \in \mathbb{G}$, the numerical method is positivity-preserving if $\mathbf{U}^{n+1} \in \mathbb{G}$. Any $\mathbf{U}^{n+1}_i \notin \mathbb{G}$ leads to an ill-posed system and the termination of the simulation.

The Lax-Friedrichs flux has the property that $\mathbf{U}^{\mathrm{LF},\pm}_i = \mathbf{U}^n_i \mp 2\lambda \hat{\mathbf{F}}^{\mathrm{LF}}_{i \pm 1/2} \in \mathbb{G}$ under the condition $\mathrm{CFL}\leq 0.5$ \cite{zhang2010positivity,zhang2011positivity,zhang2012positivity}. Therefore, positivity can be ensured by modifying the high-order numerical flux as convex combination of the original flux and the Lax-Friedrichs flux if the density or pressure is to become negative without correction \cite{hu2013positivity}. If $\rho(\mathbf{U}^+_i)<\epsilon_{\rho} = \min(10^{-13}, \rho^0_{\min})$, we compute the limiting factor by

\begin{equation}
\theta^+_{i+1/2} = \frac{\rho(\mathbf{U}^{\mathrm{LF},+}_i)-\epsilon_{\rho}}{\rho(\mathbf{U}^{\mathrm{LF},+}_i)-\rho(\mathbf{U}^+_i)} \in [0,1].
\end{equation}
If $\rho(\mathbf{U}^-_{i+1})<\epsilon_{\rho}$, the limiting factor is

\begin{equation}
\theta^-_{i+1/2} = \frac{\rho(\mathbf{U}^{\mathrm{LF},-}_{i+1})-\epsilon_{\rho}}{\rho(\mathbf{U}^{\mathrm{LF},-}_{i+1})-\rho(\mathbf{U}^-_{i+1})} \in [0,1].
\end{equation}
We modify the numerical flux by

\begin{equation}
\hat{\mathbf{F}}^*_{i+1/2} = (1-\theta_{\rho, i+1/2})\hat{\mathbf{F}}^{\mathrm{LF}}_{i+1/2}+\theta_{\rho, i+1/2}\hat{\mathbf{F}}_{i+1/2},
\end{equation}
which guarantees positive density, $\rho \left(  \mathbf{U}^{n+1}_i \right) = \rho \left( \mathbf{U}^n_i + \lambda \left( \hat{\mathbf{F}}^{*}_{i-1/2} -\hat{\mathbf{F}}^{*}_{i+1/2} \right)\right) > 0$ \cite{hu2013positivity}, where $\theta_{\rho, i+1/2} = \min(\theta^+_{i+1/2}, \theta^-_{i+1/2})$. 

Given positive density, positive pressure is enforced by limiting the flux $\hat{\mathbf{F}}^*_{i+1/2}$. If $p(\mathbf{U}^+_i)<\epsilon_{p} = \min(10^{-13}, p^0_{\min})$, the limiting factor is determined by
\begin{equation}
\theta^+_{i+1/2} = \frac{p(\mathbf{U}^{\mathrm{LF},+}_i)-\epsilon_{p}}{p(\mathbf{U}^{\mathrm{LF},+}_i)-p(\mathbf{U}^+_i)}  \in [0,1].
\end{equation}
And if $p(\mathbf{U}^-_i)<\epsilon_{p}$, the limiting factor is
\begin{equation}
\theta^-_{i+1/2} = \frac{p(\mathbf{U}^{\mathrm{LF},-}_i)-\epsilon_{p}}{p(\mathbf{U}^{\mathrm{LF},-}_i)-p(\mathbf{U}^-_i)}  \in [0,1].
\end{equation}
$\hat{\mathbf{F}}^*_{i+1/2}$ is replaced by

\begin{equation} \label{flux_limiter2}
\hat{\mathbf{F}}^{**}_{i+1/2} = (1-\theta_{p, i+1/2}\theta_{\rho, i+1/2})\hat{\mathbf{F}}^{\mathrm{LF}}_{i+1/2} + \theta_{p, i+1/2}\theta_{\rho, i+1/2} \hat{\mathbf{F}}^*_{i+1/2},
\end{equation}
where $\theta_{p, i+1/2} = \min(\theta^+_{i+1/2}, \theta^-_{i+1/2})$. This treatment ensures positive pressure $p\left( \mathbf{U}^{n+1}_i\right)  = p\left(\mathbf{U}^n_i + \lambda \left( \hat{\mathbf{F}}^{**}_{i-1/2} -\hat{\mathbf{F}}^{**}_{i+1/2} \right) \right)  > 0$.
Thus, $ \mathbf{U}^{n+1}_i \in \mathbb{G}$ if $\mathbf{U}^n \in \mathbb{G}$ under the condition $\mathrm{CFL} \leqslant 0.5$ \cite{hu2013positivity}.

The 2D extension of Eq. (\ref{governdis}) is
\begin{eqnarray}\label{governdis2d}
\mathbf{U}^{n+1}_{i,j} &=& \frac{\alpha}{2} \left(\mathbf{U}^{n}_{i,j}  + 2 \lambda_x  \hat{\mathbf{F}}_{i-1/2,j}\right) + \frac{\alpha}{2} \left( \mathbf{U}^{n}_{i,j}  - 2 \lambda_x \hat{\mathbf{F}}_{i+1/2,j} \right) \\
\nonumber
&+& \frac{1-\alpha}{2} \left(\mathbf{U}^{n}_{i,j}  + 2 \lambda_y \hat{\mathbf{F}}_{i,j-1/2}\right) + \frac{1-\alpha}{2} \left( \mathbf{U}^{n}_{i,j} - 2 \lambda_y \hat{\mathbf{F}}_{i,j+1/2} \right),
\end{eqnarray}
where $\lambda_x = \Delta t/\Delta x \alpha$ and $\lambda_y = \Delta t/\Delta y \alpha$. Following Ref. \cite{hu2013positivity}, $\alpha$ is defined as
\begin{eqnarray}
\alpha = \frac{\tau_x}{\tau_x+\tau_y}, \quad \tau_x = \frac{\Vert|u|+c\Vert_{\infty}}{\Delta x}, \quad \tau_y = \frac{\Vert|v|+c\Vert_{\infty}}{\Delta y}.
\end{eqnarray}
One can apply the positivity-preserving flux limiters in a dimension-by-dimension manner.
\subsection{MR representations}\label{sec:mr}
To achieve high computational efficiency and low memory storage the space-time adaptivity strategy developed in Ref. \cite{han2014adaptive} is used. Specifically, the MR method \cite{harten1995multiresolution} is used for mesh refinement due to its high data compression rate. 
Let $\ell$ be the integer index of levels where a smaller $\ell$ corresponds to a coarser resolution. For simplicity, the 1D conservative projection and prediction operators \cite{roussel2003conservative}, respectively, are written as
\begin{equation}
P^{\ell+1}_{\ell}(\mathbf{U}_{\ell+1}): \quad \mathbf{U}_{\ell,i} = \frac{1}{2} (\mathbf{U}_{\ell+1,2i}+\mathbf{U}_{\ell+1,2i+1}),
\end{equation}
and

\begin{eqnarray} \label{predicton}
P^{\ell}_{\ell+1}(\mathbf{U}_{\ell}): \quad \hat{\mathbf{U}}_{\ell+1,2i} &=& \mathbf{U}_{\ell,i} + \sum_{m=1}^{r}\beta_m (\mathbf{U}_{\ell,i+m}+\mathbf{U}_{\ell,i-m}),
\\ \nonumber
\hat{\mathbf{U}}_{\ell+1,2i+1} &=& \mathbf{U}_{\ell,i} - \sum_{m=1}^{r}\beta_m (\mathbf{U}_{\ell,i+m}+\mathbf{U}_{\ell,i-m}),
\end{eqnarray}
where $\beta_m$ is the interpolation coefficient of the $(2r+1)$-th order prediction. Notice that the prediction operator $P^{\ell}_{\ell+1}$ is used to predict data at $\ell+1$ by interpolating data at $\ell$. Mesh refinement and coarsening are triggered by comparing the prediction error $\mathbf{D}_{\ell,i} = \mathbf{U}_{\ell,i} - \hat{\mathbf{U}}_{\ell,i}$ with a level-dependent threshold $\epsilon_{\ell} = 2^{d (\ell -\ell_{\mathrm{max}})} \epsilon$, where $\epsilon$ is a user-defined parameter, $d$ is the space dimension and $\ell_{\mathrm{max}}$ is the maximum level of the adaptive data structure.
\section{Numerical method} \label{sec:num}
We first show that the original operators in the MR method may lead to positivity failure and can be modified to have the positivity-preserving property. Then, we discuss the generation of negative states during a LTS cycle due to fixed hierarchical time steps and the conservation flux correction and, as a remedy, we propose a modified LTS which dynamically adjusts the time steps at all different levels.
\subsection{The positivity of MR representations} \label{sec:mrpp}
\begin{propo}
For the projection operator $\mathbf{P}^{\ell+1}_{\ell} \in \mathbb{G}$ holds, while for the prediction operator $\mathbf{P}^{\ell}_{\ell+1}  \in \mathbb{G}$ may not hold.
\end{propo}
\textit{Proof}. Suppose $\mathbf{U}_{\ell+1} \in \mathbb{G}$, the projection operator $\mathbf{P}^{\ell+1}_{\ell}$ is positivity preserving as it is a convex combination of $\mathbf{U}_{\ell+1}$. $\mathbf{P}^{\ell}_{\ell+1}$ does not necessary have this property as it is not a convex combination of $\mathbf{U}_{\ell} \in \mathbb{G}$. $\square$

In order to guarantee positivity of $\mathbf{P}^{\ell}_{\ell+1}$ and as we realize that the first order prediction operator is positivity preserving, the original high-order projection operator is modified by a convex combination of itself and the first order operator. Similarly as with the positivity-preserving flux limiter, we first enforce the positivity of density. In 1D, supposing $\mathbf{U}_{\ell} \in \mathbb{G}$, if $\rho(\hat{\mathbf{U}}_{\ell+1,2i}) < \epsilon_{\rho}$ or $\rho(\hat{\mathbf{U}}_{\ell+1,2i+1}) < \epsilon_{\rho}$, the limiting factors are computed as

\begin{equation}
\theta^0_{\rho} = \frac{\rho(\mathbf{U}_{\ell,i})-\epsilon_{\rho}}{\rho(\mathbf{U}_{\ell,i}) - \rho(\hat{\mathbf{U}}_{\ell+1,2i})}  \quad \mathrm{and} \quad \theta^1_{\rho} = \frac{\rho(\mathbf{U}_{\ell,i})-\epsilon_{\rho}}{\rho(\mathbf{U}_{\ell,i}) - \rho(\hat{\mathbf{U}}_{\ell+1,2i+1})}, 
\end{equation}
respectively. With $\theta_{\rho} = \min (\theta^0_{\rho}, \theta^1_{\rho})$, the predicted values in Eq. (\ref{predicton}) are modified as

\begin{equation} \label{MR-rho}
\hat{\mathbf{U}}^*_{\ell+1,2i} = (1-\theta_{\rho}) \hat{\mathbf{U}}_{\ell+1,2i} + \theta_{\rho} \mathbf{U}_{\ell,i} \quad \mathrm{and} \quad
\hat{\mathbf{U}}^*_{\ell+1,2i+1} = (1-\theta_{\rho}) \hat{\mathbf{U}}_{\ell+1,2i+1} + \theta_{\rho} \mathbf{U}_{\ell,i}.
\end{equation}
Subsequently, we ensure positivity of pressure by

\begin{equation}\label{MR-p}
\hat{\mathbf{U}}^{**}_{\ell+1,2i} = (1-\theta_{p}) \hat{\mathbf{U}}^*_{\ell+1,2i} + \theta_{p} \mathbf{U}_{\ell,i} \quad \mathrm{and} \quad
\hat{\mathbf{U}}^{**}_{\ell+1,2i+1} = (1-\theta_{p}) \hat{\mathbf{U}}^*_{\ell+1,2i+1} + \theta_{p} \mathbf{U}_{\ell,i},
\end{equation}
where $\theta_{p} = \min (\theta^0_p, \theta^1_p)$. If $p(\hat{\mathbf{U}}_{\ell+1,2i}) < \epsilon_{p}$ or $p(\hat{\mathbf{U}}_{\ell+1,2i+1}) < \epsilon_{p}$, the corresponding factors are

\begin{equation}
\theta^0_p = \frac{p(\mathbf{U}_{\ell,i})-\epsilon_{p}}{p(\mathbf{U}_{\ell,i}) - p(\hat{\mathbf{U}}_{\ell+1,2i})}  \quad \mathrm{and} \quad \theta^1_p = \frac{p(\mathbf{U}_{\ell,i})-\epsilon_{p}}{p(\mathbf{U}_{\ell,i}) - p(\hat{\mathbf{U}}_{\ell+1,2i+1})}.
\end{equation}
\begin{thm}
The modified prediction operator $\mathbf{P}^{\ell,**}_{\ell+1}$ is positivity and conservation preserving.
\end{thm}
\textit{Proof}. Similarly as with proofs for the flux limiter, we have

\begin{equation}
\rho(\hat{\mathbf{U}}^{**}_{\ell+1,2i}) = \frac{\theta^0_p-\theta}{\theta^0_p}\rho(\mathbf{U}_{\ell,i}) + \frac{\theta}{\theta^0_p} \epsilon_{\rho}> 0
\end{equation}
and

\begin{equation}
p(\hat{\mathbf{U}}^{**}_{\ell+1,2i}) \geqslant \frac{\theta^0_p-\theta}{\theta^0_p} p(\mathbf{U}_{\ell,i}) + \frac{\theta}{\theta^0_p} \epsilon_{p} > 0,
\end{equation}
as $\mathbf{U}_{\ell,i} \in \mathbb{G}$ and $\theta = \theta_{\rho} \theta_p \leqslant \theta^0_{\rho}$, which implies $\hat{\mathbf{U}}^{**}_{\ell+1,2i} \in \mathbb{G}$. 
The conservation of this operator is easily verified due to the convex weighting form of 
Eqs. (\ref{MR-rho}) and (\ref{MR-p}). $\square$

Note that this limiter, like that in Ref. \cite{hu2013positivity}, does not affect the formal accuracy. Let $\hat{\mathbf{U}}^{\text{lim}}_{\ell+1, 2i}$ be the value after limiting, i.e. $\hat{\mathbf{U}}^{*}_{\ell+1, 2i}$ or $\hat{\mathbf{U}}^{**}_{\ell+1, 2i}$. The difference between the original predicted value $\hat{\mathbf{U}}_{\ell+1, 2i}$ and limited value $\hat{\mathbf{U}}^{\text{lim}}_{\ell+1, 2i}$ is

\begin{equation}
\Vert \hat{\mathbf{U}}_{\ell+1, 2i}- \hat{\mathbf{U}}^{\text{lim}}_{\ell+1, 2i}\Vert = (1-\theta_g) \Vert \hat{\mathbf{U}}_{\ell+1, 2i} - \mathbf{U}_{\ell, i}\Vert.
\end{equation}
As $\hat{\mathbf{U}}_{\ell+1, 2i}$ and $\mathbf{U}_{\ell, i}$ are bounded in smooth regions, the accuracy is not affected if we can show that

\begin{equation}\label{proofaccu}
1-\theta_g = \frac{\epsilon_g - g(\hat{\mathbf{U}}_{\ell+1, 2i})}{g(\mathbf{U}_{\ell, i})-g(\hat{\mathbf{U}}_{\ell+1, 2i})} \leqslant \frac{\vert \epsilon_g - g(\hat{\mathbf{U}}_{\ell+1, 2i})\vert}{g(\mathbf{U}_{\ell, i})-\epsilon_g} = O(\Delta x^k).
\end{equation}
Similar with Ref. \cite{hu2013positivity}, a sufficient condition is $\vert \epsilon_g - g(\hat{\mathbf{U}}_{\ell+1, 2i})\vert = O(\Delta x^k)$ and $g(\mathbf{U}_{\ell, i})-\epsilon_g$ is bounded away from zero. Following Ref. \cite{zhang2010positivity, hu2013positivity}, the exact solution $\mathbf{U}(x)$ is assumed to be smooth and positive (density and pressure), and gives the cell-average or nodal representation of $\tilde{\mathbf{U}}_i$ satisfying $g(\tilde{\mathbf{U}}_i) \geqslant M > 0$. Given a sufficiently small $\Delta x$, the numerical solution $\mathbf{U}_{\ell, i}$ obtained from an $p$th-order approximation satisfies

\begin{equation}
g(\mathbf{U}_{\ell, i}) - \epsilon_g \geqslant g(\tilde{\mathbf{U}}_i) - O(\Delta x^{p}) - \epsilon_g \geqslant M/2 - \epsilon_g > 0.
\end{equation}
Also we can obtain

\begin{equation}
\vert \epsilon_g - g(\hat{\mathbf{U}}_{\ell+1, 2i})\vert < \vert g(\tilde{\mathbf{U}}_{\ell+1, 2i}) - g(\hat{\mathbf{U}}_{\ell+1, 2i}) \vert = O(\Delta x^k), 
\end{equation}
as $g(\tilde{\mathbf{U}}_{\ell+1, 2i}) \geqslant M $ and $g(\hat{\mathbf{U}}_{\ell+1, 2i}) \leqslant \epsilon_g$, where $k$ is the order of the interpolation method. This completes the proof of Eq. (\ref{proofaccu}).

\subsection{The positivity of local time stepping} \label{sec:lts}
\begin{figure}[p]
\begin{center}
\includegraphics[width=1.0\textwidth]{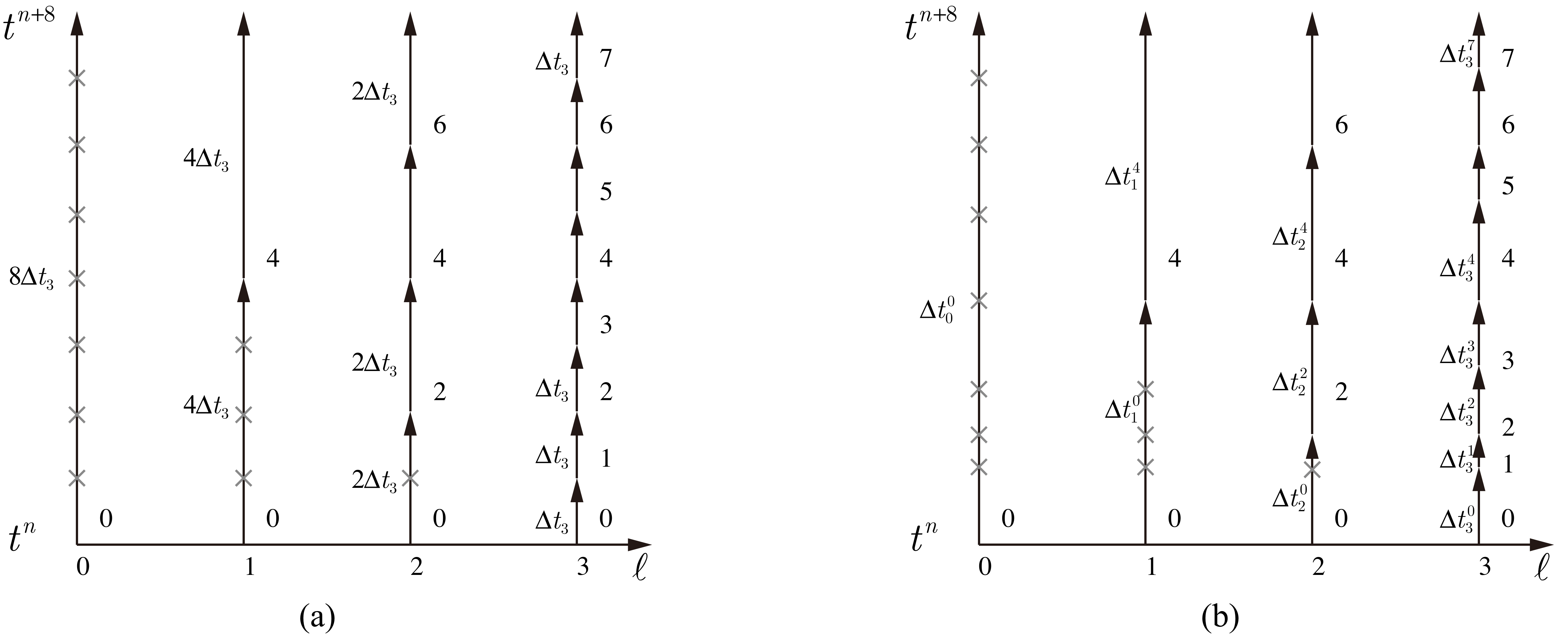}
\caption{Schematic of time cycles of LTS in Ref. \cite{domingues2008adaptive} (a) and Sec. \ref{sec:lts} (b) with $\ell_{\max} = 3$ and Euler forward time integration. Every line with arrow represets an evolution at different levels. The symbol ``$\times$'' refers to the interpolation of the intermediate states.}\label{fig_LTS}
\label{3d_basic}
\end{center}
\end{figure}

A LTS uses large time steps to evolve large scales and small time steps for fine scales, which are represented by coarse and fine grid resolutions in a MR framework, respectively. For example, the LTS developed in Ref. \cite{domingues2008adaptive} and employed in Ref. \cite{han2014adaptive} uses $2^{\ell_{\max}-\ell} \Delta t_{\ell_{\max}}$ for different levels ($0 \leqslant \ell \leqslant \ell_{\max}$) in the MR representation, where $\Delta t_{\ell_{\max}}$ is the time step for the finest level

\begin{equation}
\Delta t_{\ell_{\max}} = \frac{\mathrm{CFL}\, \Delta x_{\ell_{\max}}}{\Vert|u|+c\Vert^{n}_{\infty}},
\end{equation}
with $\Delta x_{\ell_{\max}}$ being the cell size at the finest level and $\Vert|u|+c\Vert^{n}_{\infty}$ computed at $t^n$. The superscript ``n'' is the timestep index during a LTS cycle.
Thus during a full LTS time cycle, the solutions are advanced from $t^n$ to $t^n+2^{\ell_{\max}} \Delta t_{\ell_{\max}}$ within $2^{\ell_{\max}}$ substeps, as shown in Fig. \ref{fig_LTS}(a). Note that the time step at each level is fixed during the entire cycle. Despite its simplicity, this scheme exhibits positivity failure during a full cycle wherein $\text{CFL} \leqslant 0.5$ may be invalid when the actual $\Vert|u|+c\Vert_{\infty}$ is larger than $\Vert|u|+c\Vert^{n}_{\infty}$ especially for large $\ell_{\max}$. As a consequence, we compute the time step of $\ell_{\max}$ at every substep of the cycle. For simplicity, we consider the Euler forward time integration to describe the basic idea of our LTS which can be easily extended to multi-stage Runge-Kutta schemes \cite{shu1988efficient}. For the example $\ell_{\max} = 3$, a full LTS time cycle is sketched in Fig. \ref{fig_LTS}(b). First we need to determine

\begin{equation}\label{dt}
\Delta t^m_{\ell_{\max}} = \frac{\mathrm{CFL}\, \Delta x_{\ell_{\max}}}{\Vert|u|+c\Vert^{m}_{\infty}}, \quad  0 \leq m \leq 2^{\ell_{\max}}-1
\end{equation}
at $\ell=\ell_{\max}$, where the superscript ``m'' is the index of timestep during the LTS cycle. The time steps at the coarse levels are calculated subsequently from

\begin{equation}
\Delta t^m_{\ell} = \Delta t^{m}_{\ell+1} + \Delta t^{m+2^{\ell_{\max}-1-\ell}}_{\ell+1}.
\end{equation}
To make sure that $\lambda^m_{\ell} a^m_0 \leqslant 0.5$ at every level, with $\lambda^m_{\ell} = \Delta t^m_{\ell}/\Delta x_{\ell}$ and $a^m_0 = \Vert|u|+c\Vert^m_{\infty}$, we limit the $\Delta t^{m+2^{\ell_{\max}-1-\ell}}_{\ell+1}$ by reassigning

\begin{equation}\label{dtlimit}
\Delta t^{m+2^{\ell_{\max}-1-\ell}}_{\ell+1} \leftarrow \min(\Delta t^{m}_{\ell+1}, \Delta t^{m+2^{\ell_{\max}-1-\ell}}_{\ell+1}),
\end{equation}
as

\begin{equation}\label{dtlimitproof}
\frac{\Delta t^m_{\ell}}{\Delta x_{\ell}} a^m_0 = 0.5 \left[ \frac{\Delta t^m_{\ell+1}}{\Delta x_{\ell+1}} + \frac{\Delta t^{m+2^{\ell_{\max}-1-\ell}}_{\ell+1}}{\Delta x_{\ell+1}} \right] a^m_0 \leqslant \frac{\Delta t^m_{\ell+1}}{\Delta x_{\ell+1}} a^m_0 \leqslant 0.5.
\end{equation}
Then the flow fields are advanced by the Euler forward scheme as example for a Runge-Kutta sub-step,

\begin{equation}
\mathbf{U}^{m+2^{\ell_{\max}-\ell}}_{i} = \mathbf{U}^{m}_{i} + \frac{\Delta t^m_{\ell}}{\Delta x_{\ell}} \left( \hat{\mathbf{F}}^{m}_{\ell,i-1/2} -\hat{\mathbf{F}}^{m}_{\ell,i+1/2} \right),
\end{equation}
according to the sequence in Fig. \ref{fig_LTS}, i.e., the evolution at $\ell$ is performed only when two evolution steps at $\ell+1$ are completed. 

For example, consider $\mathbf{U}^n \in\mathbb{G}$ in Fig. \ref{fig_LTS}(b) where $\ell_{\max}=3$ and evolve level $2$ for $m=0$ and level $3$ for $m=0,1$. According to Eq. (\ref{dt}), $\mathbf{U}^{0,\ell=3} \in\mathbb{G}$ after advection by $\Delta t^0_3$. We update the primitive value based on $\mathbf{U}^{0,\ell=3}$ and calculate the timestep $\Delta t^1_3$ by Eq. (\ref{dt}). Then the timestep constraint leads to $\Delta t^{1}_{3} = \min(\Delta t^{0}_{3}, \Delta t^{1}_{3})$. Thus we can evolve level $3$ by $\Delta t^{1}_{3}$ and level $2$ by $\Delta t^{0}_{2} = \Delta t^{0}_{3} + \Delta t^{1}_{3}$. Both evolved values, $\mathbf{U}^{1,\ell=3}$ and $\mathbf{U}^{0,\ell=2}$, are in $\mathbb{G}$, due to Eq. (\ref{dtlimitproof}).
Therefore, positivity is maintained as $\mathbf{U}^{\mathrm{LF}, \pm}_i = \mathbf{U}_i \mp 2\lambda \hat{\mathbf{F}}^{\mathrm{LF}}_{i \pm 1/2} \in \mathbb{G}$ holds during the full LTS cycle. 

As shown in Fig. \ref{fig_LTS}(b), the intermediate states are obtained by interpolation at $\ell < \ell_{\max}$ when the finer level $\ell+1$ requires a prediction from $\ell$ to update its block boundary value,

\begin{equation}
\mathbf{U}^{*,m+2^{\ell_{\max}-\ell}} = \left( 1-\frac{\Delta t^{m}_{\ell}}{t^{m}_{\ell}} \right) \mathbf{U}^{m} + \frac{\Delta t^{m}_{\ell}}{t^{m}_{\ell}} \mathbf{U}^{m+2^{\ell_{\max}-\ell}},
\end{equation}
where the accumulated time is $t^{m}_{\ell} = \sum_m \Delta t^{m}_{\ell}$. It is also positivity preserving as $\frac{\Delta t^{m}_{\ell}}{t^{m}_{\ell}} \in [0,1]$ and $\mathbf{U}^{m}, \mathbf{U}^{m+2^{\ell_{\max}-\ell}} \in \mathbb{G}$. 
As mentioned in Ref. \cite{domingues2008adaptive}, this treatment limits the time integration scheme to 2nd-order Runge-Kutta methods.

To maintain strict conservation a conservative flux correction \cite{domingues2008adaptive} is adopted between cells with different levels. For instance, if the cell size at the left side of the interface is $\Delta x^{\ell+1}$ and size of the right side is $\Delta x^{\ell}$, the left most cell of $\ell$ is updated by

\begin{equation}
\mathbf{U}^{m}_{\ell} = \mathbf{U}^{m}_{\ell} - \frac{\Delta t^m_{\ell}}{\Delta x_{\ell}} \hat{\mathbf{F}}^{m}_{\ell, \mathrm{L}} - \frac{1}{2}\left( \frac{\Delta t^m_{\ell+1}}{\Delta x_{\ell+1}} \hat{\mathbf{F}}^{m}_{\ell+1, \mathrm{R}} + \frac{\Delta t^{m+2^{\ell_{\max}-\ell}}_{\ell+1}}{\Delta x_{\ell+1}} \hat{\mathbf{F}}^{m+2^{\ell_{\max}-\ell}}_{\ell+1, \mathrm{R}}\right).
\end{equation}
\begin{rmk}
It is not mandatory to apply the flux limiter to the ghost cells for a single block domain, i.e., $i=-1$ and $i=N+1$, where $N$ is the number of cells in x direction. In a MR grid, we do need apply the flux limiter at the block interface and the limiter should be applied to the coarser cell if the two blocks have different cell sizes. 
\end{rmk}
\begin{thm}
The LTS is positivity preserving after the conservative flux correction.
\end{thm}
\textit{Proof}. Similarly as with proofs for the flux limiter, we suppose that the level at the left side of the interface is $\ell+1$, and at the right side it is $\ell$. So the two cells at each side of the interface are updated by

\begin{eqnarray}\label{eq_flux_corr_1}
\mathbf{U}^{n+1}_{\ell+1, N} = \mathbf{U}^{n}_{\ell+1, N} + \frac{\Delta t^0}{\Delta x} \left( \hat{\mathbf{F}}^{**,0}_{\ell+1, N-1/2} -\hat{\mathbf{F}}^{**,0}_{\ell+1, N+1/2} \right) \\
\nonumber
\mathbf{U}^{n+2}_{\ell+1, N} = \mathbf{U}^{n+1}_{\ell+1, N} + \frac{\Delta t^1}{\Delta x} \left( \hat{\mathbf{F}}^{**,1}_{\ell+1, N-1/2} -\hat{\mathbf{F}}^{**,1}_{\ell+1, N+1/2} \right)
\end{eqnarray}
and

\begin{equation}\label{eq_flux_corr_2}
\mathbf{U}^{n+2}_{\ell, 0} = \mathbf{U}^{n}_{\ell, 0} + \frac{\Delta t^0 + \Delta t^1}{2\, \Delta x}  \left( \hat{\mathbf{F}}^{**}_{\ell, -1/2} -\hat{\mathbf{F}}^{**}_{\ell, 1/2} \right),
\end{equation}
respectively. After the conservative flux correction, Eq. (\ref{eq_flux_corr_2}) becomes

\begin{equation}\label{eq_flux_corr_3}
\mathbf{U}^{n+2}_{\ell, 0} = \mathbf{U}^{n}_{\ell, 0} + \frac{1}{2} \left( \frac{\Delta t^0}{\Delta x} \hat{\mathbf{F}}^{**,0}_{\ell+1, N-1/2} + \frac{\Delta t^1}{\Delta x} \hat{\mathbf{F}}^{**,1}_{\ell+1, N-1/2} \right) - \frac{\Delta t^0 + \Delta t^1}{2\, \Delta x} \hat{\mathbf{F}}^{**}_{\ell, 1/2}.
\end{equation}
Clearly, Eq. (\ref{eq_flux_corr_1}) is positivity preserving if $\mathrm{CFL}\leq 0.5$, as Eq. (\ref{eq_flux_corr_3}) can be rewritten as

\begin{eqnarray}\label{eq_flux_corr_4}
\mathbf{U}^{n+2}_{\ell, 0} &=& \frac{1}{4} \left( \mathbf{U}^{n}_{\ell, 0} + 2  \frac{\Delta t^0}{\Delta x} \hat{\mathbf{F}}^{**,0}_{\ell+1, N-1/2} \right) + \frac{1}{4} \left( \mathbf{U}^{n}_{\ell, 0} + 2  \frac{\Delta t^1}{\Delta x} \hat{\mathbf{F}}^{**,1}_{\ell+1, N-1/2} \right) \\
\nonumber
&+& \frac{1}{2} \left( \mathbf{U}^{n}_{\ell, 0} -  \frac{\Delta t^0 + \Delta t^1}{2\, \Delta x} \hat{\mathbf{F}}^{**}_{\ell, 1/2}\right).
\end{eqnarray}
The first and second terms are in $\mathbb{G}$ due to the positivity-preserving flux limiter while the third term is in $\mathbb{G}$ due to Eq. (\ref{dtlimit}). Thus $\mathbf{U}^{n+2}_{\ell, 0} \in \mathbb{G}$, as it is a convex combination of three elements in $\mathbb{G}$. $\square$

\subsection{Accuracy test} \label{sec:accuracy}
The main objective of MR method is to achieve high compression rate for large-scale simulations with acceptable errors rather than high asymptotic convergence rates.
Although the application of high-order scheme generally improves quality of the solution,   
due to the complex operations involved and nonlinearity of the governing equations, 
it is very hard to assess analytically whether such high formal order  can be maintained in general cases. 
However, in some simple linear cases, we observe high-order accuracy by suitably bounding the errors at the coarser levels.

Consider that the error at the level $\ell$ is $\varepsilon^{\ell} = \Vert u^{\ell}_{e} - u^{\ell}_{\text{MR}} \Vert \leqslant \Vert u^{\ell}_{e} - u^{\ell}_{\text{num}} \Vert + \Vert u^{\ell}_{\text{num}} - u^{\ell}_{\text{MR}} \Vert$, where the subscripts `e', `num' and `MR' refer to the exact solution, the numerical solution on a uniform grid and the results after performing the MR representations, respectively. We know that the discretization error $\Vert u^{\ell}_{e} - u^{\ell}_{\text{num}} \Vert$ of a given high-order discretization scheme is $O(2^{-\ell \, k}\Delta x_{0}^k)$, where $k$ is the truncation-error order. We can bound the error generated by the MR representation, $\Vert u^{\ell}_{\text{num}} - u^{\ell}_{\text{MR}} \Vert \leqslant \varepsilon_r \Delta x_0^k$, where the reference tolerance $\varepsilon_r$ is a small constant parameter. If the prediction error at level $\ell$ exceeds $\varepsilon_r \Delta x_0^k$, this level should be refined. 

\begin{figure}[p]
\begin{center}
\hspace*{-0.8cm}\subfloat[][]{\label{accuracya}\includegraphics[scale=0.3]{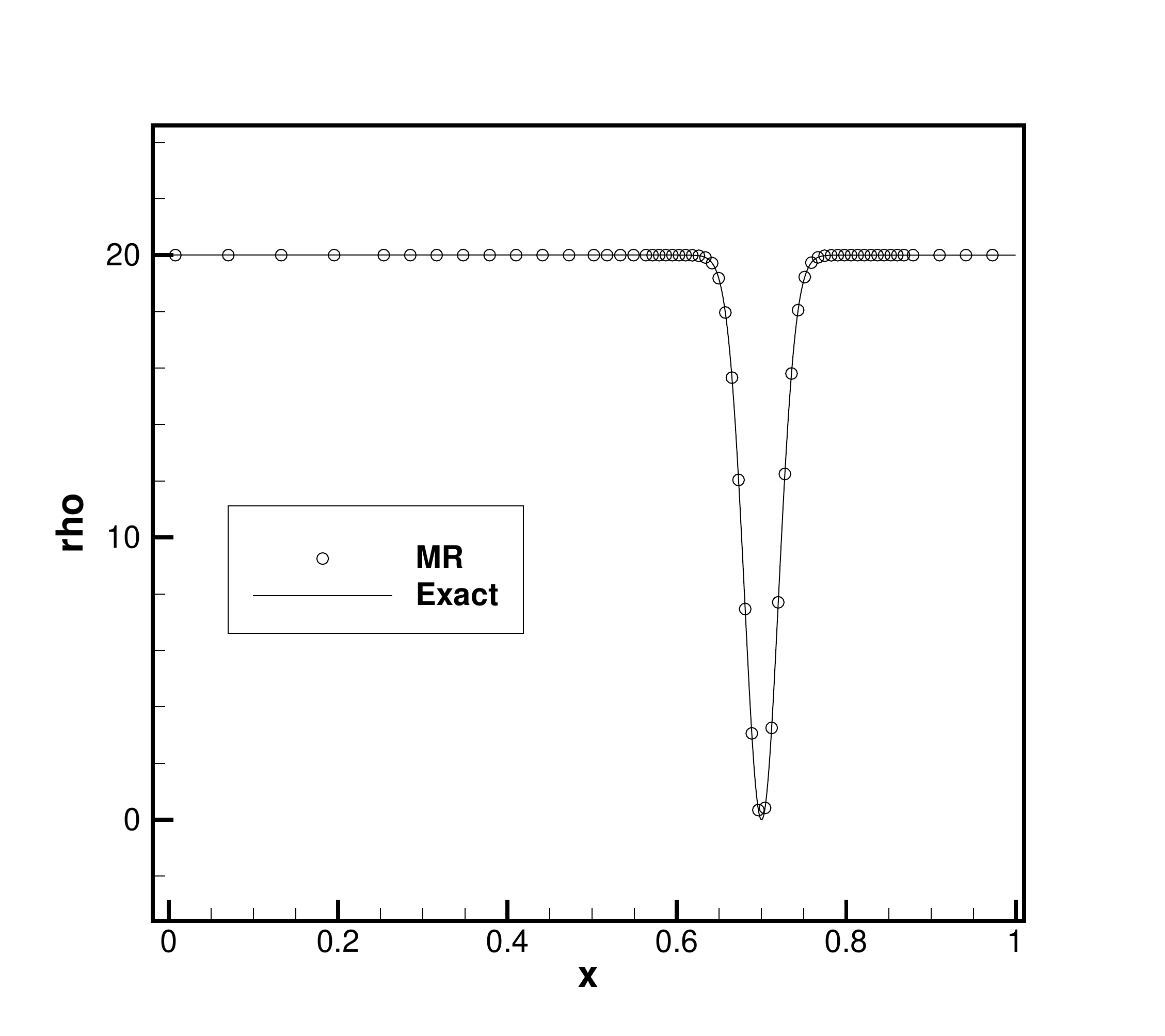}}
\subfloat[][]{\label{accuracyb}\includegraphics[scale=0.3]{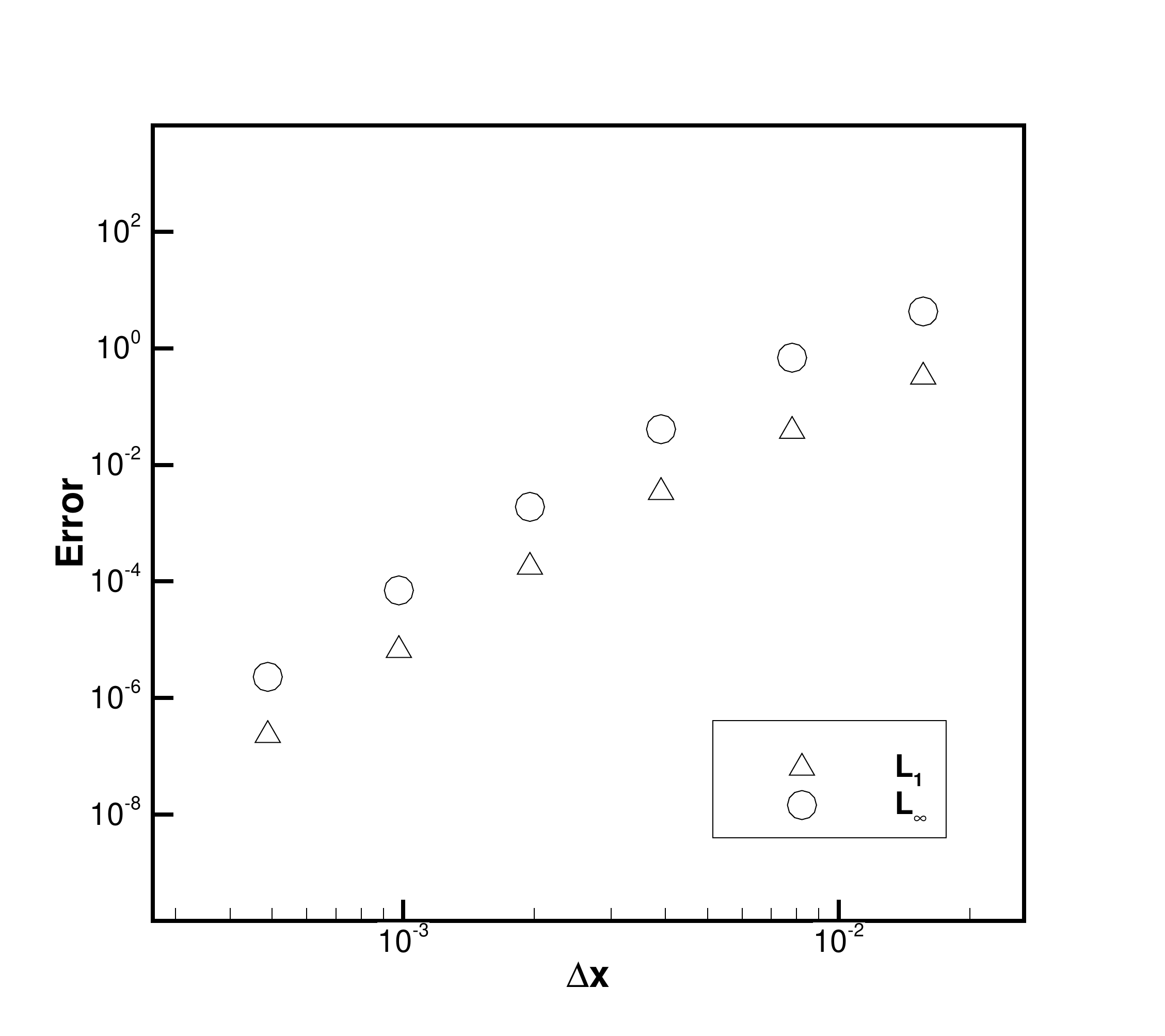}}
\caption{Linear advection problem: (a) the profile after advection with $\Delta x_{\ell_\text{max}}=\frac{1}{512}$ and (b) $L_1$ and $L_{\infty}$ error with increasing resolution. Gird points in (a) are plotted every $4$ points.}
\label{accuracy}
\end{center}
\end{figure}

To assess the accuracy of the present positivity method with suitably bounded errors, 
we consider a linear advection case with an initial Gaussian function $19.99999 [ 1- e^{-\frac{1}{2}\left( \frac{x-0.5}{0.02} \right)^2} ]$ in the domain $[0,1]$. Periodic boundary conditions are applied at the left and right sides of the domain. A 5th-order WENO scheme and 2nd-order Runge-Kutta scheme are used. Here, the time-step size $\Delta t = \Delta x^{5/3}$ is used to keep the spatial errors dominant. The accuracy test is performed by decreasing the grid size at all levels with $\ell_{\max}=4$ and $\varepsilon_r = 0.01$. As shown in Fig. \ref{accuracya}, the MR results indicate that the grid is only refined to the finest level near the corner of the Gaussian function. The $L_1$ and $L_{\infty}$ norms measured at $t=0.2$ in Fig. \ref{accuracyb} show that the expected high-order accuracy is achieved.

Indeed, the order of accuracy will be reduced if the chosen error tolerance is large, however, the compression rate becomes larger. There is a trade-off between accuracy and compression rate in the MR framework. Note that strict high-order accuracy may not be guaranteed in more complex cases, as the required tolerance $\varepsilon_r$ is small, which degenerates the MR method to a uniform grid method, i.e., the compression rate is $0$. 

\section{Numerical examples}\label{sec:example}

In this section, we apply our numerical method to simulate a number of 1D and 2D test cases, where high-order conservative schemes may fail. The spatial discretization scheme is the 5th-order finite difference WENO scheme and Lax-Friedrichs flux is used. The 2nd-order TVD Runge-Kutta scheme \cite{shu1988efficient} is used for time marching. If not mentioned otherwise, the CFL number is $0.5$ and $\gamma$ is $1.4$. The MR and LTS are employed for every case which previous positivity-preserving method \cite{hu2013positivity} can not pass. The parameter $\epsilon$ in the refinement threshold is $0.01$.

\subsection{One-dimensional cases}
\begin{figure}[p]
\begin{center}
\hspace*{-0.8cm}\subfloat[][]{\label{double_rarea}\includegraphics[scale=0.3]{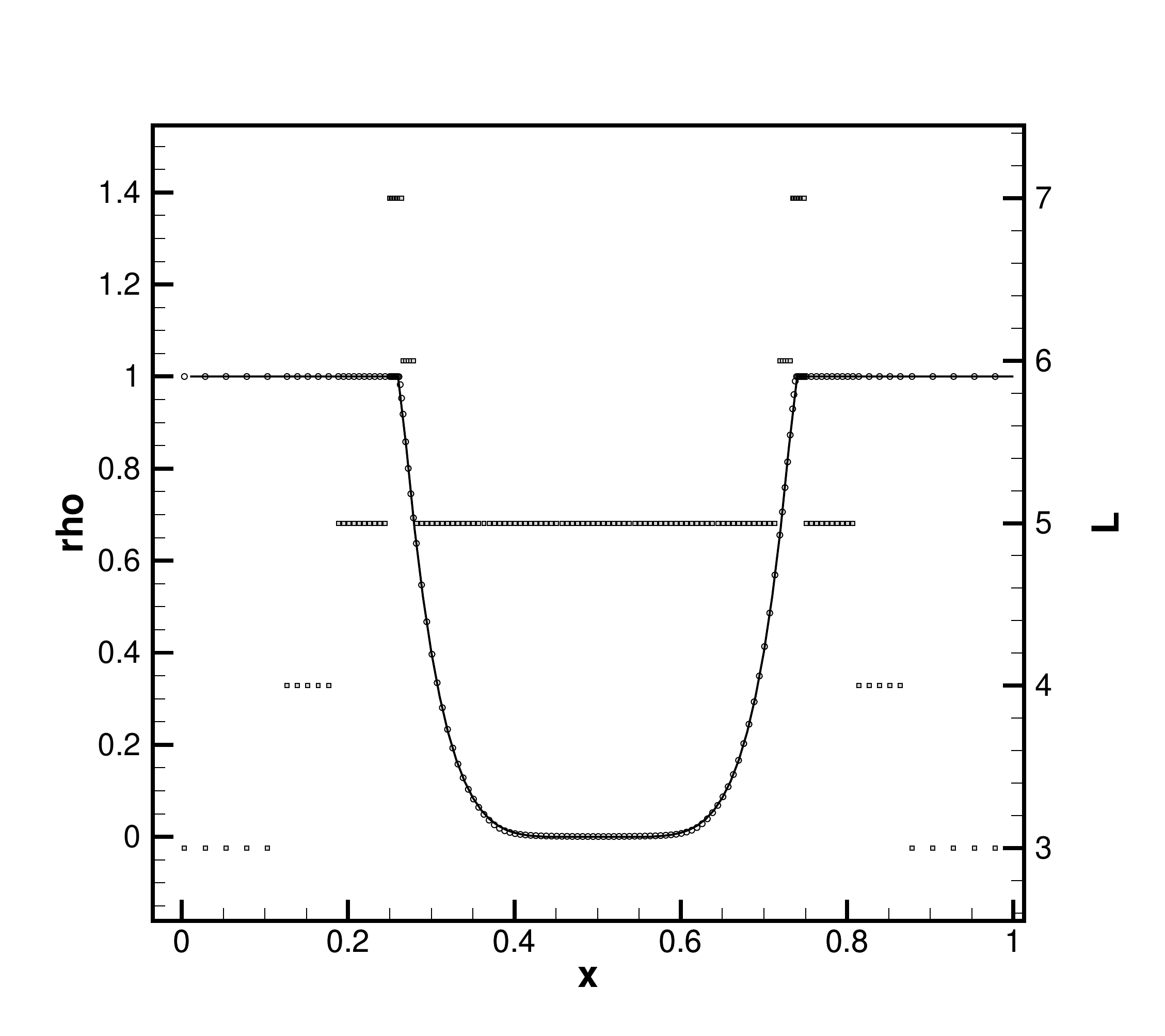}}
\subfloat[][]{\label{double_rareb}\includegraphics[scale=0.3]{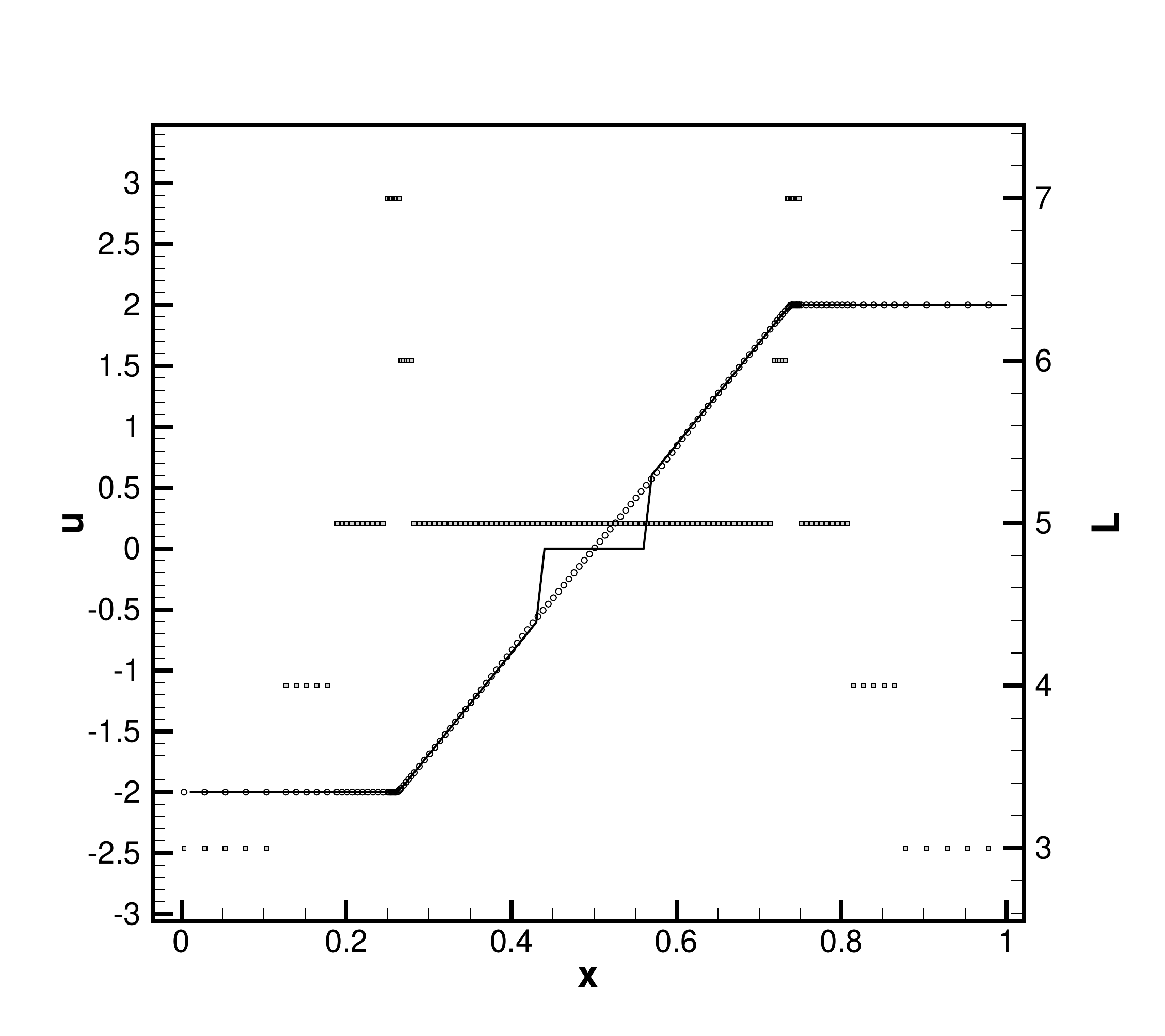}}
\caption{The MR simulation of the double-rarefaction problem: (a) density and (b) velocity profiles.}
\label{double_rare}
\end{center}
\end{figure}

\begin{figure}[p]
\begin{center}
\hspace*{-0.8cm}\subfloat[][]{\label{twoblasta}\includegraphics[scale=0.3]{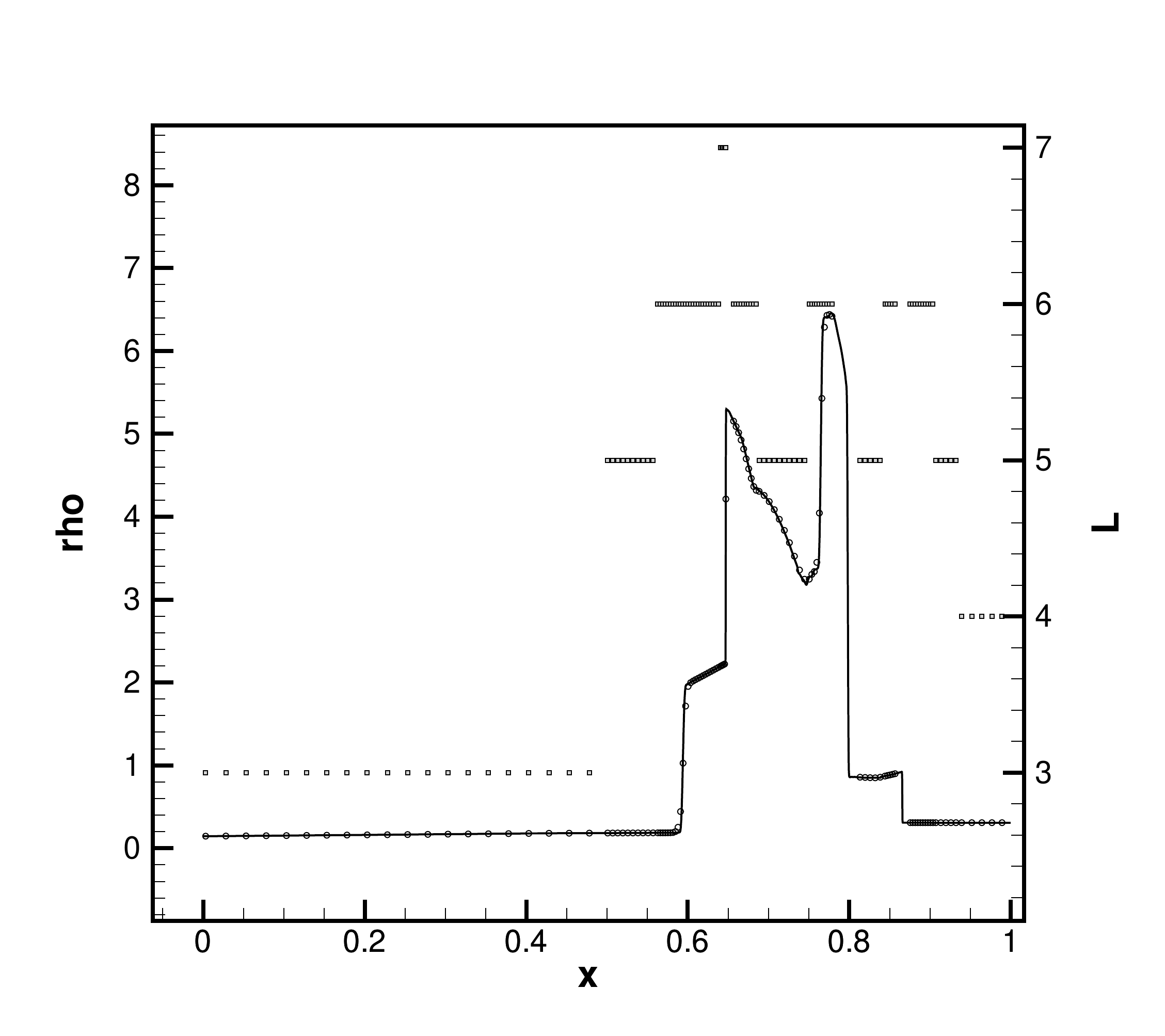}}
\subfloat[][]{\label{twoblastb}\includegraphics[scale=0.3]{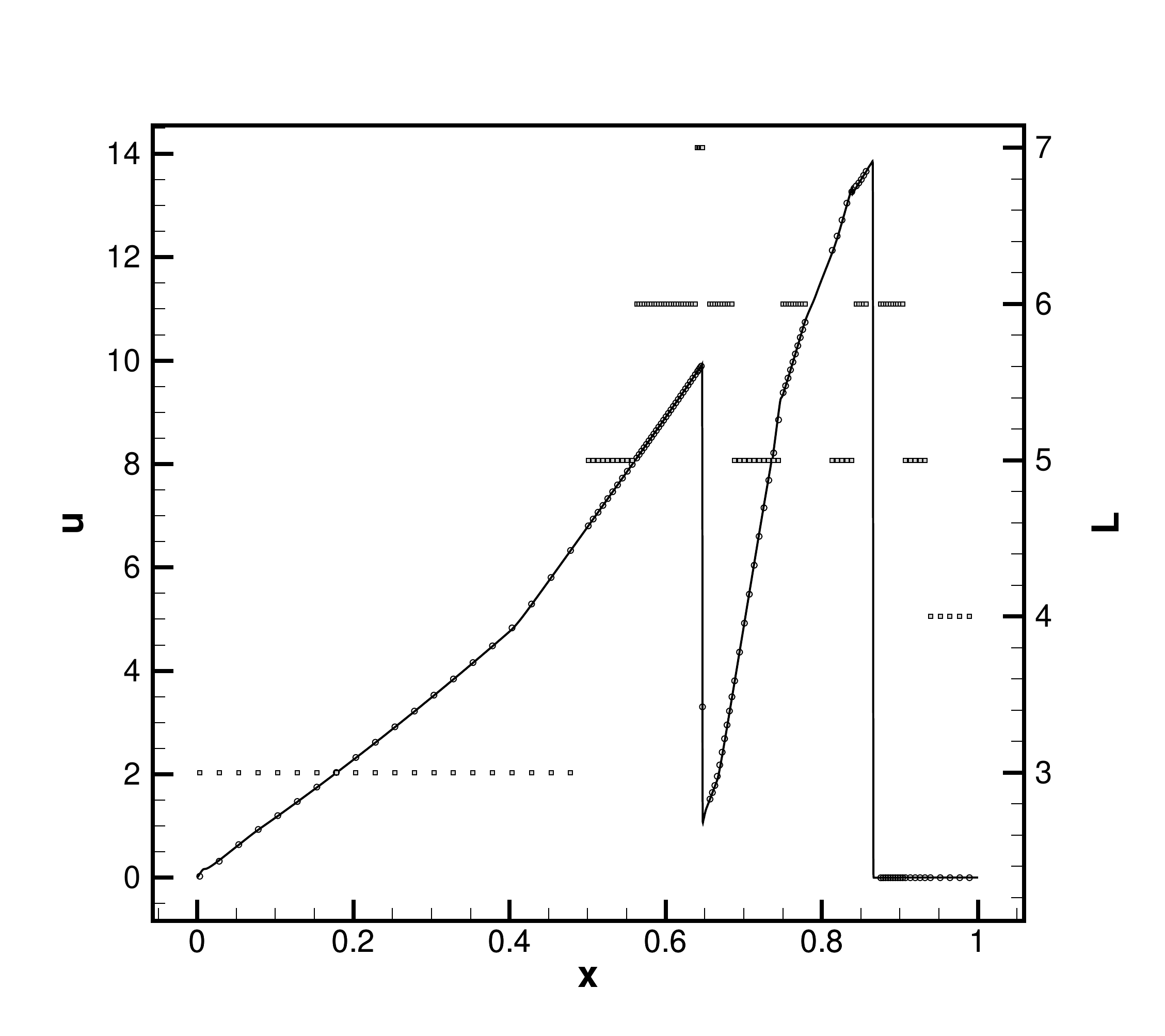}}
\caption{The MR simulation of the two blast-wave problem: (a) density and (b) velocity profiles.}
\label{twoblast}
\end{center}
\end{figure}

\begin{figure}[p]
\begin{center}
\hspace*{-0.8cm}\subfloat[][]{\label{LeBlanca}\includegraphics[scale=0.3]{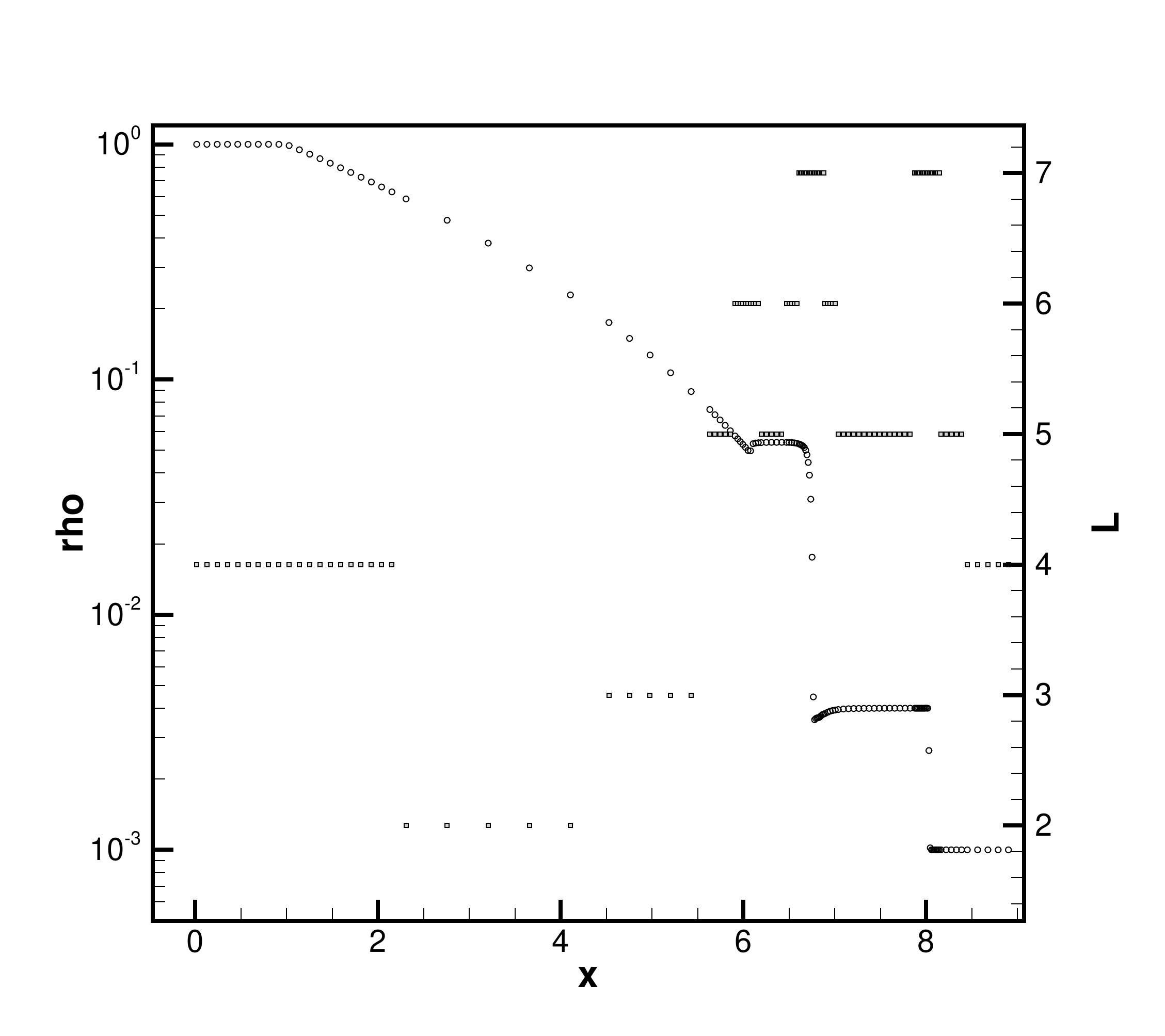}}
\subfloat[][]{\label{LeBlancb}\includegraphics[scale=0.3]{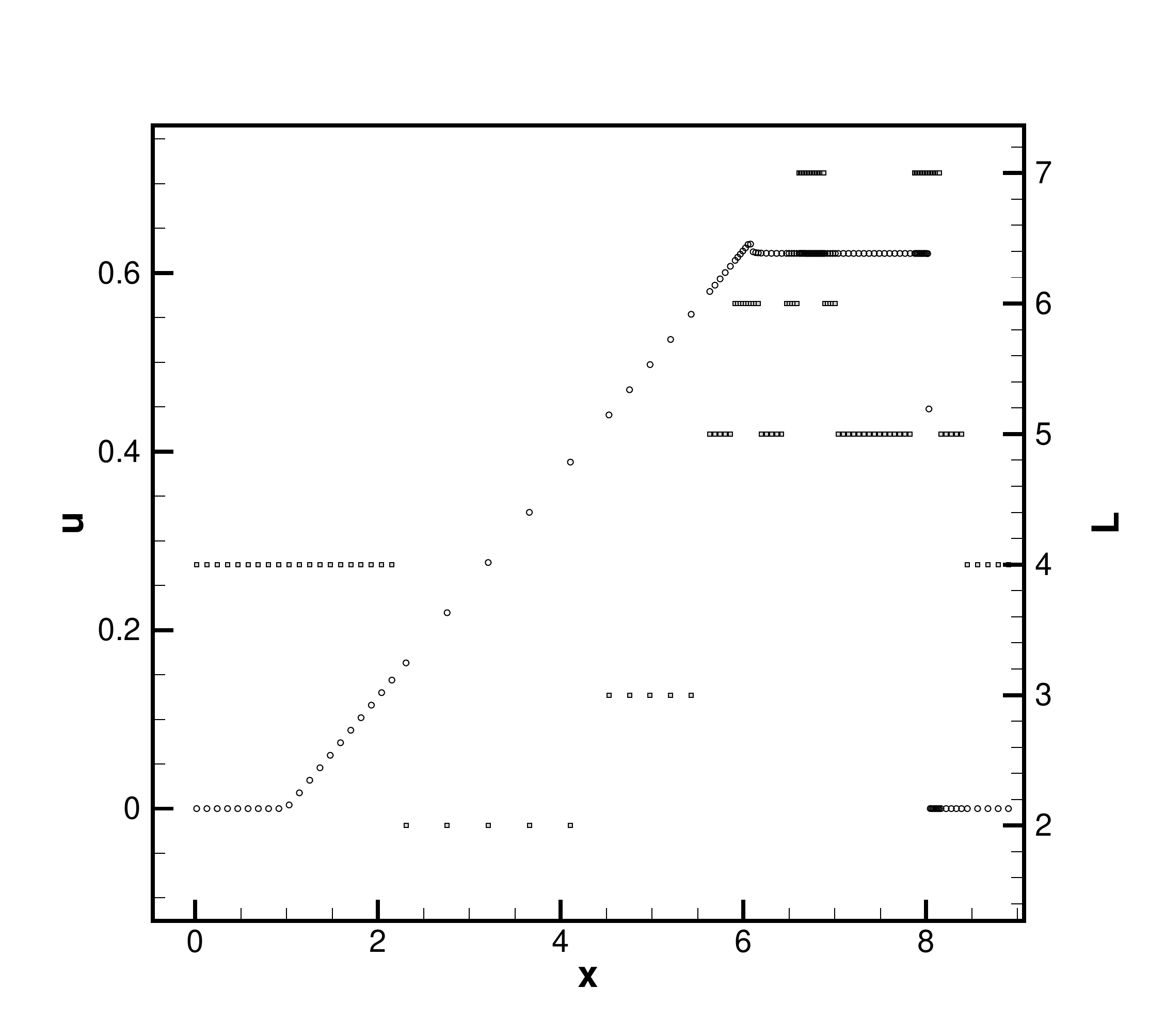}}
\caption{The MR simulation of the Le Blanc shock-tube problem: (a) density and (b) velocity profiles.}
\label{LeBlanc}
\end{center}
\end{figure}

Three 1D cases, either one containing vacuum or strong discontinuities, are considered. The first case is the double-rarefaction problem where vacuum occurs \cite{hu2013positivity}. The initial condition is 

\begin{eqnarray}
\left( \rho,u,p \right) = 
\begin{cases}
\left( 1, -2, 0.1\right) & \quad \text{if $0<x<0.5$}\\
\left( 1,  2, 0.1\right) & \quad \text{if $0.5<x<1$}
\end{cases}.
\end{eqnarray}
There is one block at the coarsest level and the maximum level is $\ell_{\max} = 7$, with each block containing $20$ inner cells. The final time is $t=0.1$. Fig. \ref{double_rare} shows the density and velocity profiles (symbol ``$\circ$'') which exhibit good agreement with exact solution (solid lines). The vacuum region is accurately captured by the density profile. Note that the symbols are plotted every $4$ points to show highly non-uniform distributed cells, i.e., only cells near discontinuities are refined. The corresponding value of $\ell$ of every cell is plotted by ``{\tiny$\square$}''. The second 1D case is the two blast-wave interaction problem \cite{woodward1984numerical} which contains strong discontinuities and has the initial condition

\begin{eqnarray}
\left( \rho,u,p \right) = 
\begin{cases}
\left( 1, 0, 10^3\right) & \quad \text{if $0<x<0.1$}\\
\left( 1, 0, 10^{-2}\right) & \quad \text{if $0.1<x<0.9$}\\
\left( 1, 0, 10^{2}\right) & \quad \text{if $0.9<x<1$}
\end{cases}.
\end{eqnarray}
Simulations are performed with one block at the coarsest level and $\ell_{\max} = 7$ till $t=0.038$. Reflective conditions are employed at the left and right boundaries. The density and velocity distributions are exactly the same with the reference solution which is a high-resolution numerical result calculated in Ref. \cite{hu2013positivity}, as shown in Fig. \ref{twoblast}. High resolution blocks only appear in very few regions, which indicates that much less cells are needed to achieve a similar result with the reference solution. The intial condition of the Le Blanc problem \cite{loubere2005subcell, zhang2010positivity, hu2013positivity} is

\begin{eqnarray}
\left( \rho,u,p \right) = 
\begin{cases}
\left( 1, 0, \frac{2}{3}\times 10^{-1}\right) & \quad \text{if $0<x<3$}\\
\left( 10^{-3}, 0, \frac{2}{3}\times 10^{-10}\right) & \quad \text{if $3<x<9$}
\end{cases}.
\end{eqnarray}
We refine one block at the coarsest level to $\ell_{\max} = 7$. The final time is $t=6$. A good agreement with the exact solution is observed in Fig. \ref{LeBlanc}. One can notice that the cell distribution is very sparse. Cells which are refined to $\ell_{\max}$ only exist near the shock and contact discontinuity.

\subsection{Two-dimensional cases}
\begin{figure}[p]
\begin{center}
\includegraphics[width=1.0\textwidth]{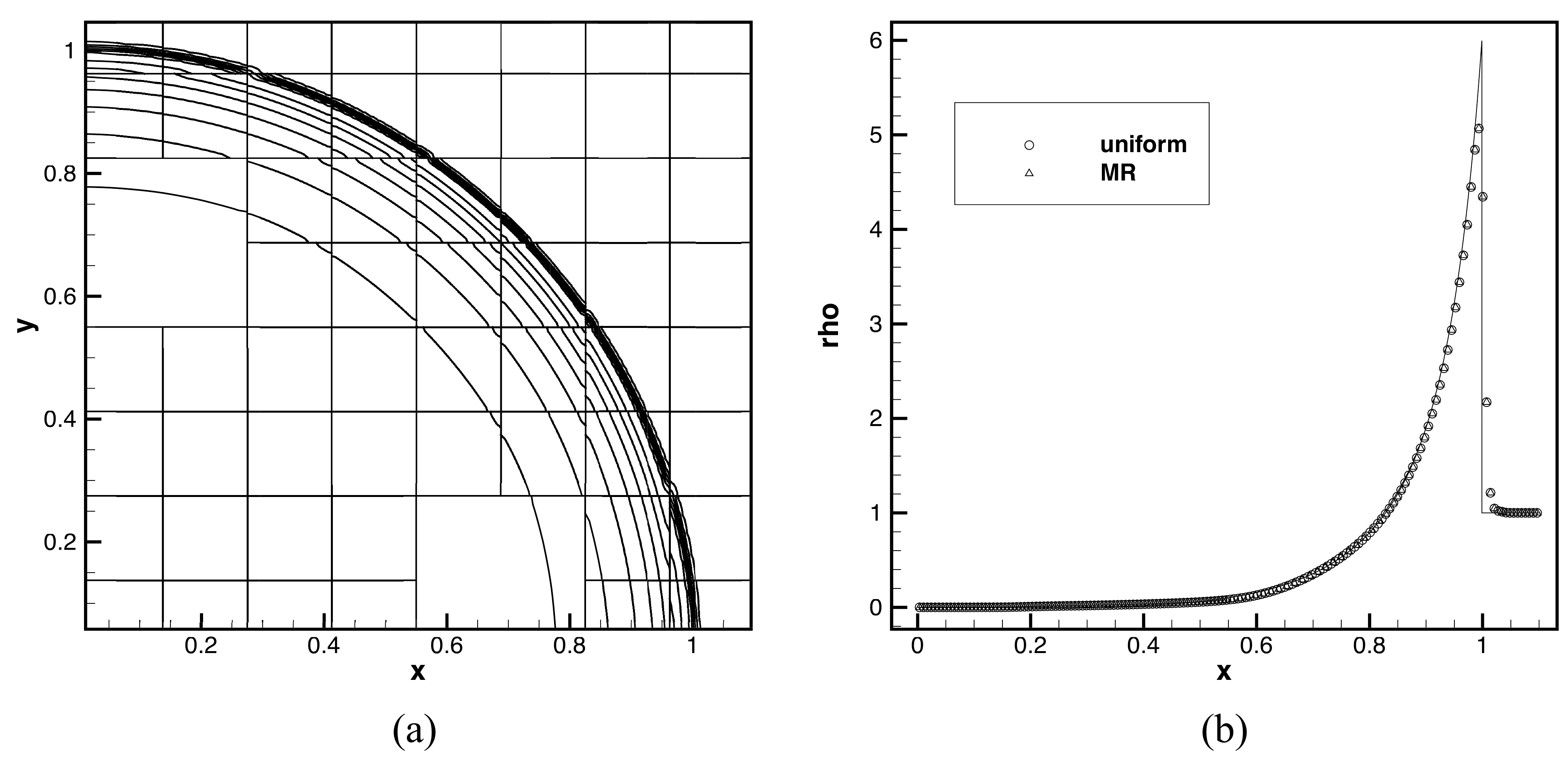}
\caption{Two-dimensional Sedov problem: (a) $10$ density contours from $0$ to $6$; (b) density profile along $y=0$.}
\label{sedov}
\end{center}
\end{figure}

We consider two 2D cases in Ref. \cite{zhang2010positivity, hu2013positivity} for comparison. The first one is the two-dimensional Sedov problem \cite{zhang2010positivity, hu2013positivity}. The computational domain is $[0,1.1] \times [0,1.1]$, where the lower-left corner cell has high energy,

\begin{eqnarray}
\left( \rho,u,v,p \right) = 
\begin{cases}
\left( 1, 0, 0, 4\times 10^{-13}\right) & \quad \text{if $x>\Delta x$, $y>\Delta y$}\\
\left( 1, 0, 0, \frac{9.79264}{\Delta x \Delta y}\times 10^{4}\right) & \quad \text{else}
\end{cases}.
\end{eqnarray}
The coarsest level has one block and are refined to the $\ell_{\max} = 3$. The final time is $t=10^{-3}$. Reflective boundary conditions are employed at the lower and left boundaries, and outflow conditions are employed at the right and upper boundaries. The MR simulation result plotted in Fig. \ref{sedov} is comparable to those in Refs. \cite{zhang2010positivity, hu2013positivity}. And the density profile along $y=0$ of MR results agrees the uniform grid result and the exact solution very well.

\begin{figure}[p]
\begin{center}
\includegraphics[width=0.8\textwidth]{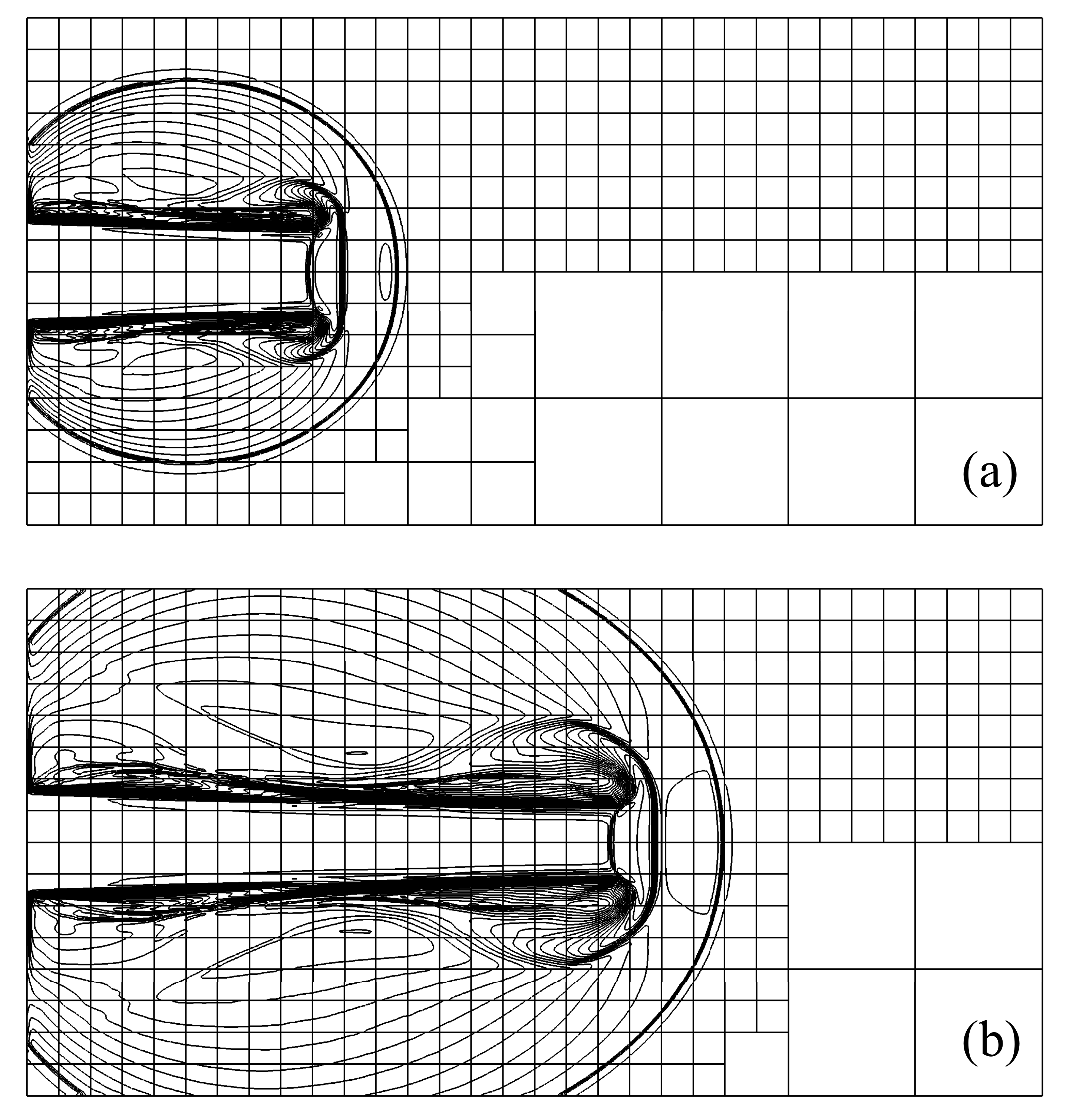}
\caption{Comparison of density contours for uniform mesh (upper) and MR (lower) simulations of the Mach-2000 jet problem at (a) $t=5.0 \times 10^{-4}$ and (b) $t = 1.0 \times 10^{-3}$. Logarithmic scales from $-4$ to $4$.}
\label{mach2000}
\end{center}
\end{figure}

\begin{figure}[p]
\begin{center}
\includegraphics[width=1.0\textwidth]{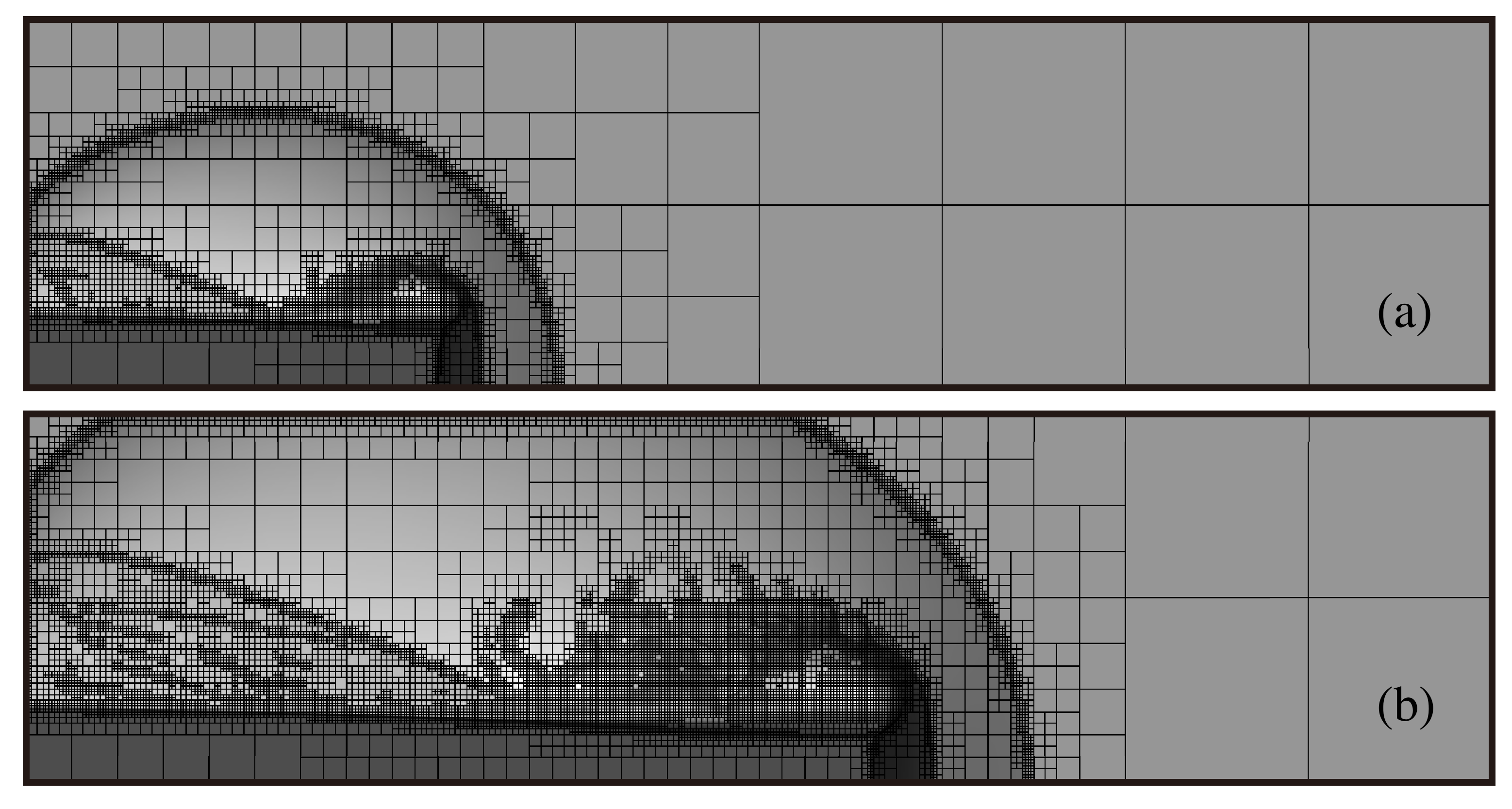}
\caption{Density contours and MR representations of Mach-2000 jet problem at (a) $t=5.0 \times 10^{-4}$ and (b) $t = 1.0 \times 10^{-3}$. Logarithmic scales from $-4$ to $4$.}
\label{mach2000mesh}
\end{center}
\end{figure}

\begin{figure}[p]
\begin{center}
\hspace*{-1.5cm}\includegraphics[width=15cm]{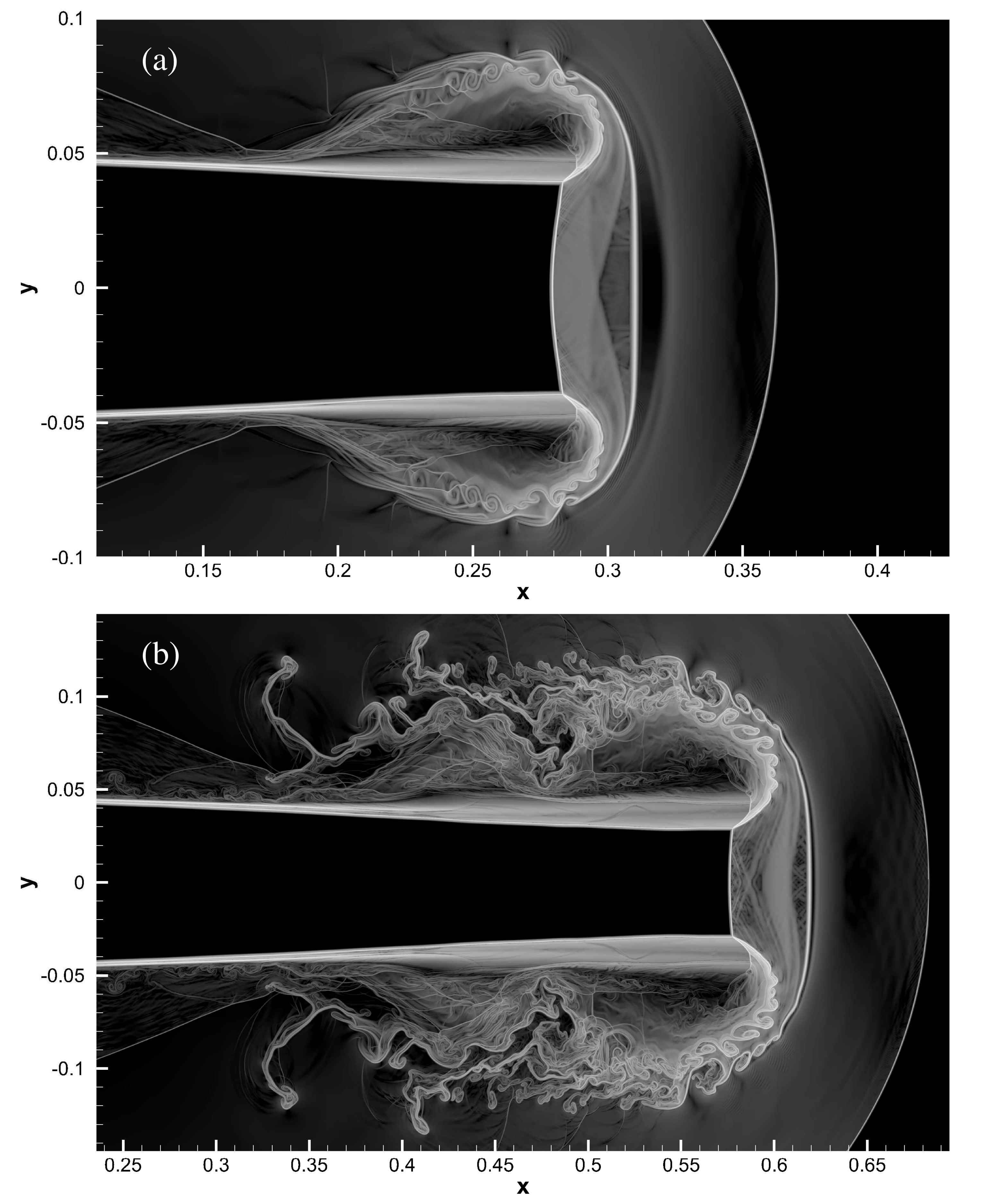}
\caption{Density gradient contours of Mach-2000 jet problem at (a) $t=5.0 \times 10^{-4}$ and (b) $t = 1.0 \times 10^{-3}$. Logarithmic scales from $0$ to $12$.}
\label{mach2000dr}
\end{center}
\end{figure}

The Mach-2000 jet problem studied in Refs \cite{zhang2010positivity, zhang2011positivity, zhang2012positivity} is considered here. The computational domain is $[0,1] \times [0,0.25]$ which is initialized uniformly with $\left(\rho,u,v,p \right) = \left( 0.5,0,0,0.4127\right)$. Symmetry conditions are applied at the lower boundary, an outflow condition is applied at the right and upper boundaries, and an inflow condition with states

\begin{eqnarray}
\left( \rho,u,v,p \right) = 
\begin{cases}
\left( 5, 800, 0, 0.4127\right) & \quad \text{if $y<0.05$}\\
\left( 0.5, 0, 0, 0.4127\right) & \quad \text{else}
\end{cases}
\end{eqnarray}
is applied at the left boundary. The CFL number is $0.25$, the final time is $t=10^{-3}$ and $\gamma = 5/3$. Simulations are performed with $4 \times 1$ blocks at the coarsest level and $\ell_{\max}=3$, leading to an effective resolution of $640 \times 160$ at the finest level. For comparison, uniform mesh numerical simulation is also performed. As shown in Fig. \ref{mach2000} the difference between the uniform mesh and MR numerical result is minor. Also note that the numerical results are in good agreement with previous result in Ref. \cite{hu2013positivity}. We also conduct a MR simulation with $\ell_{\max}=7$ (effective resolution is $10240 \times 2560$) to test our numerical method in a high-resolution adaptive mesh. The density contours and MR representations are shown in Fig. \ref{mach2000mesh} at $t=5.0 \times 10^{-4}$ and $t = 1.0 \times 10^{-3}$. The block distribution is highly sparse and blocks are only refined to $\ell_{\max}=7$ near shock waves, shear layer and small structures. The density gradient contours in Fig. \ref{mach2000dr} show small vortical features due to shear layer instabilities near the top region of the jet.

\section{Concluding remarks} \label{sec:con}
In this paper we have proposed a positivity-preserving method for MR simulations of compressible flows involving extreme conditions such as near vacuum states and strong discontinuities. The main contribution is to modify two steps of the MR method which lead to positivity failure. First, by limiting the high-order interpolated values we construct the prediction operator which is positivity and conservation preserving. Second, a LTS which dynamically adjusts the time steps at all different levels addresses the positivity failure. Also we provide proof that positivity is strictly preserved for every internal step of a LTS cycle, and that the conservation flux correction is positivity preserving under a time step constraint.
A number of 1D and 2D test cases are used to demonstrate that the positivity-preserving property is successfully achieved. This method has the potential to be applied in MR simulations of more complex flows such as magnetohydrodynamics and multiphase flows.
\section*{Acknowledgment}
This work is supported by China Scholarship Council under No. 201306290030,
National Natural Science Foundation of China (No. 11628206) and 
Deutsche Forschungsgemeinschaft (HU 1527/6-1).
\bibliographystyle{plain}
\bibliography{aipsamp}

\end{document}